\documentclass[structabstract]{aa}

\usepackage{url}
\usepackage[breaklinks=true]{hyperref}
\usepackage{twoopt}
\usepackage[english]{babel}          
\usepackage{pdflscape}

\usepackage{natbib}
\bibpunct{(}{)}{;}{a}{}{,} 

\usepackage[utf8]{inputenc}
\usepackage[normalem]{ulem}
\usepackage{siunitx}
\usepackage{amsmath, amssymb}
\usepackage[farskip=0pt]{subfig}

\usepackage{multirow,bigstrut,ctable}
\usepackage{rotating}
\usepackage[varg]{txfonts} 
\usepackage{threeparttable}

\def \deg         {\text{$^{\circ}$}}
\def \arcmin      {\text{$^\prime$}}

\begin{document}

\title{The effect of the ionosphere\\ on ultra-low frequency radio-interferometric observations}

\titlerunning{The ionosphere at low-frequencies}

\author{F. de Gasperin\inst{1,2} \and M. Mevius\inst{3} \and D.A. Rafferty\inst{2} \and H.T. Intema\inst{1} \and R.A. Fallows\inst{3}}

\authorrunning{F.~de~Gasperin et al.}

\institute{
Leiden Observatory, Leiden University, P.O.Box 9513, NL-2300 RA, Leiden, The Netherlands, \email{fdg@strw.leidenuniv.nl}
\and Hamburger Sternwarte, Universit\"at Hamburg, Gojenbergsweg 112, D-21029, Hamburg, Germany
\and ASTRON - the Netherlands Institute for Radio Astronomy, P.O.Box 2, NL-7990 AA, Dwingeloo, the Netherlands
}

\date{Received ... / Accepted ...}

\abstract
{The ionosphere is the main driver of a series of systematic effects that limit our ability to explore the low frequency ($<1$ GHz) sky with radio interferometers. Its effects become increasingly important towards lower frequencies and are particularly hard to calibrate in the low signal-to-noise ratio regime in which low-frequency telescopes operate.}
{In this paper we characterize and quantify the effect of ionospheric-induced systematic errors on astronomical interferometric radio observations at ultra-low frequencies ($<100$ MHz). We also provide guidelines for observations and data reduction at these frequencies with the Low Frequency Array (LOFAR) and future instruments such as the Square Kilometre Array (SKA).}
{We derive the expected systematic error induced by the ionosphere. We compare our predictions with data from the Low Band Antenna (LBA) system of LOFAR.}
{We show that we can isolate the ionospheric effect in LOFAR LBA data and that our results are compatible with satellite measurements, providing an independent way to measure the ionospheric total electron content (TEC). We show how the ionosphere also corrupts the correlated amplitudes through scintillations. We report values of the ionospheric structure function in line with the literature.}
{The systematic errors on the phases of LOFAR LBA data can be accurately modelled as a sum of four effects (clock, ionosphere 1st, 2nd, and 3rd order). This greatly reduces the number of required calibration parameters, and therefore enables new efficient calibration strategies.}


\maketitle

\section{Introduction}

The ultra-low frequencies ($10-100$~MHz) are the last poorly explored window available for ground-based astronomical observations. Some attempts have been made in the past to cover this frequency range, notably the 38 MHz 8th Cambridge survey \citep[8C;][]{Rees1990} and the 74 MHz Very Large Array Low-frequency Sky Survey \citep[VLSS;][]{Cohen2007,Lane2014} pioneered this exploration. More recently the GaLactic and Extragalactic All-sky MWA Survey \citep[GLEAM;][]{Hurley-Walker2017} produced images down to 72 MHz. A major limitation when observing at these frequencies is the presence of the ionosphere, a layer of partially ionised plasma, surrounding our planet.

The ionisation of the ionosphere is driven by the UV and X-ray radiation generated by the Sun during the day and is balanced by recombination at night. A lower level of ionisation is maintained during the night by the action of cosmic rays. The peak of the free electron density lies at a height of $\sim300$~km but the ionosphere extends approximately from 75 to 1000 km. The free electron column density along a line of sight (LoS) through the ionosphere is generally referred to as the total electron content (TEC). The TEC unit (TECU) is $10^{16}$~m$^{-2}$, which is the order of magnitude typically observed at zenith during nighttime; in normal conditions the TEC during daytime is ten times higher. When observing radio emission at long wavelengths, the ionosphere introduces systematic effects such as reflection, refraction, and propagation delay of the radio waves \citep{Mangum2014}. For interferometric observation, propagation delay is the main concern \citep{Intema2009}. The effect is caused by a varying refractive index $n$ of the ionospheric plasma along the wave trajectories. The total propagation delay, integrated along the LoS at frequency $\nu$, results in a phase rotation given by

\begin{equation}\label{eq:delay}
 \Phi_{\rm ion} = - \frac{2\pi\nu}{c} \int_{\rm LoS} \left( n - 1 \right)\ {\rm d}l.
\end{equation}

An $n$ constant in time and space would impose a coherent phase error that would result in a spatial shift of the observed image compared to the true sky. The problem becomes more complicated as $n$ depends strongly on time and position. 

Neglecting the frictional force and assuming a cold, collisionless, magnetised plasma (such as the ionosphere), the refractive index $n$ can be calculated exactly \citep{Davies1990}. For signals with frequencies $\nu \gg \nu_p$ (the plasma frequency, that for the ionosphere is around $1-10$~MHz), it can be expanded \citep[see e.g.][]{Datta-Barua2008} into a third-order Taylor approximation retaining only terms up to $\nu^{-4}$:

\begin{equation}\label{eq:n}
\begin{aligned}
 n \approx 1 - \frac{q^2}{8\pi^ 2m_e \epsilon_0} \cdot \frac{\textcolor{red}{n_e}}{\nu^2} \pm \frac{q^3}{16 \pi^3 m_e^2 \epsilon_0} \cdot \frac{\textcolor{red}{n_e B} \cos\theta}{\nu^3} -\\
 \frac{q^4}{128 \pi^4 m_e^2 \epsilon_0^2} \cdot \frac{\textcolor{red}{n_e^2}}{\nu^4} - \frac{q^4}{64 \pi^4 m_e^3 \epsilon_0} \cdot \frac{\textcolor{red}{n_e B^2} (1+\cos^2\theta)}{\nu^4},
\end{aligned}
\end{equation}

\begin{table*}[ht]
\centering
\begin{tabular}{lcccccc}
dTEC (TECU) & I ord & II ord (day/night) & I ord & II ord (day/night) & I ord & II ord (day/night)\\
                             & 30 MHz & 30 MHz    & 60 MHz & 60 MHz  & 150 MHz & 150 MHz\\
0.5 (remote st., bad iono.)  & 8067   & 294 / 214 & 4033   & 73 / 50 & 1613    & 12 / 8 \\
0.1 (remote st., good iono.) & 1613   & 126 / 46  & 806    & 31 / 10 & 323     & 5 / 2 \\
0.03 (across FoV)            & 404    & 97 / 16   & 242    & 24 / 4  & 96      & 4 / $<1$\\
0.01 (core st.)              & 160    & 88 / 8    & 80     & 22 / 2  & 31      & 4 / $<1$ \\
\end{tabular}
\caption{Typical ionospheric phase errors in degrees}\label{tab:err}
\end{table*}

where $n_e$ is number density of free electrons, $B$ is the magnetic field strength, $\theta$ is the angle between the magnetic field $\vec{B}$ and the electromagnetic wave propagation direction, $q$ is electron charge, $m_e$ is electron mass, and $\epsilon_0$ is electric permittivity in vacuum. In red we show the parameters related to ionospheric conditions and the Earth's magnetic field. The first term is associated with a dispersive delay proportional to the TEC along the LoS. This is the dominant term; for most radio-astronomical applications at frequencies higher than a few hundreds of MHz, higher order terms can be ignored. The second term is related to Faraday rotation, the positive sign is associated with left-hand polarised signals and the negative sign with right-hand polarised signals. This term depends on TEC and the Earth's magnetic field. The last two terms are usually ignored but can become relevant for observations at frequencies below 40~MHz. Of these last two terms, the first is dominant and depends on the spatial distribution of the electrons in the ionosphere \citep{Hoque2008}. The term becomes larger if electrons are concentrated in thin layers and not uniformly distributed.






By using Eq.~\ref{eq:n} we can give order of magnitude estimates of the expected effects at 1st and 2nd order \citep[see also][chapter 9]{Petit2010}:

\begin{equation}
 \begin{aligned}
  \delta\Phi_1 = -8067 \left( \frac{\nu}{60\ \rm MHz} \right)^{-1} \left( \frac{\rm d TEC}{1 \rm TECU} \right) [\rm deg];\\
  \delta\Phi_2 = \pm 105 \left( \frac{\nu}{60\ \rm MHz} \right)^{-2} \left[ \left( \frac{\rm d TEC}{1 \rm TECU} \right) \right. \\
        + \left. \left( \frac{\rm TEC}{1 \rm TECU} \right) \cdot \left( \frac{{\rm d}B}{40\ \mu \rm T} \right) \right] [\rm deg];
 \end{aligned}
\end{equation}

where we adopt a magnetic field $B = 40\ \mu \rm T$ with $\theta = 45\deg$. The total TEC in quiet geomagnetic conditions, can vary from $\sim1$ to $\sim20$ TECU from night to day respectively and influences the second order term. Considering a differential TEC (dTEC) of $\simeq 0.3$ TECU, which is a plausible number for baselines $\sim 50$~km (see Fig.~\ref{fig:effects}), and observing at 60~MHz, the first order term produces phase variations of several times $2\pi$. The Faraday rotation instead produces an effect of around $\pm30\deg/50\deg$ (with different signs for the two circular polarisations) at night/day assuming $\rm{d}B=1\%$. This effect is not negligible and needs to be corrected (see also Table~\ref{tab:err}). The effect quickly becomes relatively more severe at lower frequencies because of the $1/\nu^2$ dependency. Higher order effects can be mostly ignored at 60 MHz. However, at frequency $\sim 20$~MHz, the third order effect can produce large phase errors.

\subsection{LOFAR}

LOFAR \citep{vanHaarlem2013} is a radio interferometer that operates at low (including ultra-low) frequencies: $10-240$~MHz. It has 38 stations (aperture arrays capable of multi-beam forming) in the Netherlands, divided into twenty-four ``core stations'', concentrated within 4 km, and fourteen ``remote stations'', providing baselines up to $\sim120$ km\footnote{An up-to-date outline of LOFAR station positions is available at \url{http://astron.nl/lofartools/lofarmap.html}.}. The six innermost stations are packed within 1 km$^2$ and are collectively called the ``superterp''. Thirteen ``international stations'' are spread across Europe, but they are not considered in this paper. LOFAR uses two antenna types: High Band Antenna (HBA, used to observe in the frequency range $110-240$~MHz) and Low Band Antenna (LBA, used to observe in the frequency range $10-90$~MHz). In this paper we will consider only data from the LBA system, that is the most strongly affected by measurement corruptions induced by the ionosphere. However, most of the results can be extended to higher frequencies.

The LOFAR core is located at $52\deg54'32''$ N, $6\deg52'08''$ E. The telescope is marginally affected by ionospheric gradients generated at low latitude by Rayleigh-Taylor instabilities and ionospheric irregularities typical of the auroral regions. At high latitudes, strong refractive index gradients due to field-aligned ionisation structures can cause severe scintillation conditions. The ionosphere conditions in the polar regions are regularly monitored by the Kilpisj\"arvi Atmospheric Imaging Receiver Array \citep[KAIRA;][]{McKay-Bukowski2015}, a station built using LOFAR hardware in arctic Finland \citep[see e.g.][]{Fallows2016}. This said, the ionospheric impact on a specific LOFAR observation is strongly dependent on the global ionospheric conditions at the time of observation combined with the location of the target in the local sky.

\begin{figure}[t]
  \centering
  \includegraphics[width=.5\textwidth]{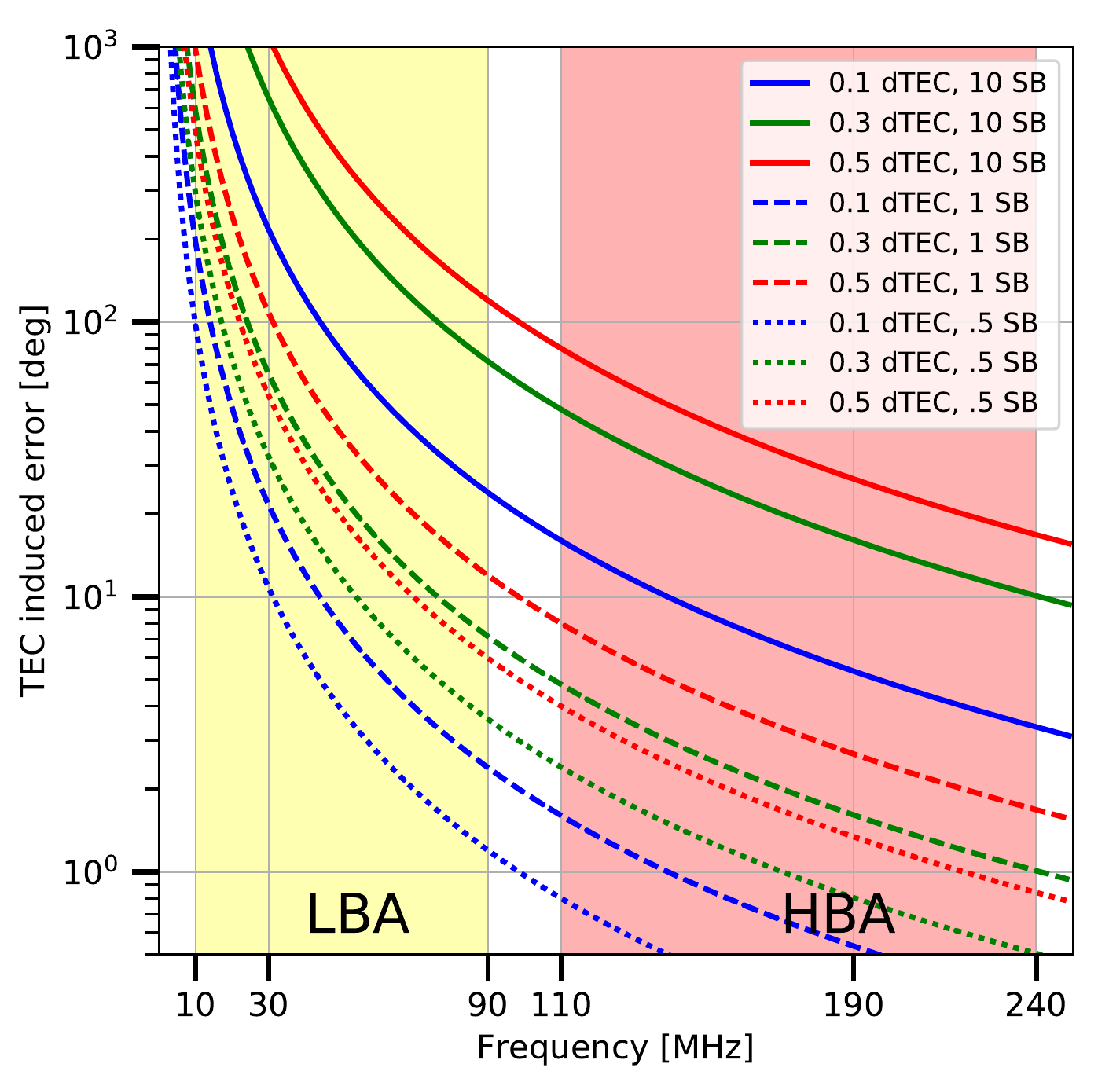}
  \caption{Ionospheric-induced phase variations between the beginning and the end of a band of 1/2, 1 and 10 LOFAR sub bands ($1 {\rm SB} \simeq 0.195$ MHz). The dTEC is assumed to be 0.1, 0.3 and 0.5 TECU. These are typical values for distances of a few tens of km. A phase variation larger than $\sim100\deg$ creates strong decorrelation. This plot can be used to estimate the maximum amount of averaging (or the maximum band usable to find a single solution) before decorrelating the signal. Coloured bands are the frequency ranges observed by LOFAR.}
  \label{fig:tecsmear}
\end{figure}

\begin{table*}
\centering
\begin{tabular}{lp{.2\textwidth}p{.2\textwidth}p{.2\textwidth}}
Target Name & 3C196 & 3C295 & 3C380 \\
RA (J2000)  & 08:13:36.1 & 14:11:20.3 &18:29:31.8 \\
Dec (J2000) & +48:13:02 & +52:12:10 & +48:44:46 \\
\hline
Date        & 03 May 2013 &
17 Nov 2017\newline
18 Nov 2017\newline
19 Nov 2017\newline
20 Nov 2017\newline
21 Nov 2017\newline
25 Nov 2017\newline
26 Nov 2017\newline
28 Nov 2017
&
17 Nov 2017\newline
18 Nov 2017\newline
19 Nov 2017\newline
20 Nov 2017\newline
21 Nov 2017\newline
24 Nov 2017\newline
25 Nov 2017\newline
26 Nov 2017\newline
28 Nov 2017\\
\hline
Time range (UTC) & 18:00 $\rightarrow$ 23:30 (5.5 hr) &
07:00 $\rightarrow$ 12:04 (5 hr)\newline
07:00 $\rightarrow$ 12:04 (5 hr)\newline
07:00 $\rightarrow$ 12:04 (5 hr)\newline
07:00 $\rightarrow$ 12:04 (5 hr)\newline
07:00 $\rightarrow$ 12:04 (5 hr)\newline
07:00 $\rightarrow$ 12:04 (5 hr)\newline
07:00 $\rightarrow$ 12:04 (5 hr)\newline
06:00 $\rightarrow$ 11:04 (5 hr)
&
12:05 $\rightarrow$ 15:07 (3 hr)\newline
12:05 $\rightarrow$ 15:07 (3 hr)\newline
12:05 $\rightarrow$ 15:07 (3 hr)\newline
12:05 $\rightarrow$ 15:07 (3 hr)\newline
12:05 $\rightarrow$ 14:06 (2 hr)\newline
14:00 $\rightarrow$ 15:00 (1 hr)\newline 
12:05 $\rightarrow$ 15:07 (3 hr)\newline
12:05 $\rightarrow$ 15:07 (3 hr)\newline
11:05 $\rightarrow$ 14:07 (3 hr)\\
\hline
Time resolution (s) & 5 & 4 & 4 \\
Frequency range (MHz) & $22-70$ & $42-66$ & $42-66$\\
Frequency resolution (kHz) & 195.3 (244 channels) & 48.8 (122 channels) & 48.8 (122 channels) \\
Recorded polarisations & XX XY YX YY  & XX XY YX YY & XX XY YX YY \\
\end{tabular}
\caption{Observation details}\label{tab:obs}
\end{table*}

\section{Data}

For this paper we analysed a set of LOFAR LBA observations pointed at three calibrators: 3C196, 3C295, and 3C380 (see Table~\ref{tab:obs}). For the detailed analysis of ionospheric systematic effects we used a 5.5 hr observation pointed at 3C196 and obtained on May 3rd, 2013 (18:00 $\rightarrow$ 23:30 UTC). The observation was taken covering a large continuous bandwidth ($22-70$ MHz) so to evaluate the effect of the ionosphere down to the lowest frequencies. Furthermore, the large bandwidth is an essential tool for separating the various effects based on their different frequency dependencies. The dataset has been calibrated using procedures that will be described in de Gasperin et al. (in preparation). Here we will focus on the outcome of the calibration to describe the influence of the ionosphere on the signal measured by the antennas. For each station pair, radio interferometers record streams of data, called visibilities. While ionospheric effects are clearly present in the visibilities, it is easier to analyse them by looking at the solutions. Solutions give a station-based representation of the systematic effects in the form of complex gain factors derived when observing a point source with a known position and flux density (i.e. a calibrator). 

We also analysed seven 8-hr long observations obtained between November 17th and November 28th, 2017. These observations were taken during day-time, pointing the LOFAR beam towards the calibrator sources 3C295 (five hrs each run) and 3C380 (three hrs each run). One observing run of 3C380 failed after two hours and was combined with an additional hour taken three days later. These observations have a narrower frequency coverage ($42-66$ MHz), centred around the bandpass peak response of the LBA dipoles ($\sim 58$ MHz). Both the frequency and time resolution of these observations are slightly higher than the one described above; however, this does not affect our results.

Because of the low antenna sensitivity and the high sky temperature, the LOFAR LBA system is often in a relatively low signal-to-noise ratio regime when observing celestial radio sources. A common way to compensate for this is to average data into larger time or frequency intervals when finding solutions. At low frequencies the optimal solution intervals are a trade-off between signal-to-noise ratio and decorrelation. The ionosphere tends to vary very quickly and averaging over more than $\sim 5$~s in time often does not allow for these changes to be tracked. Furthermore, combining too many frequency channels is also not advisable. As shown in Fig.~\ref{fig:tecsmear}, between the edges of a single LOFAR subband ($1 {\rm SB} \simeq 0.195$ MHz) centred at 30 MHz, there is a differential phase of 100\deg{} (assuming two stations with a differential TEC value of 0.5 TECU). For compact arrays (baselines shorter than a few km, with dTEC $\lesssim0.1$ TECU), this constraint can be relaxed. To facilitate accurate calibration for all baselines and ionospheric conditions, our calibration has been performed at a relatively high time and frequency resolution, as described in Table~\ref{tab:obs}.

All ionospheric-related systematic effects described in this paper induce direction dependent errors (DDE). This means that the effects vary appreciably as a function of viewing direction, even across the field of view (FoV) of the telescope. LOFAR LBA is characterised by a full width half maximum of $\sim 4\deg$. Errors across the FoV are particularly problematic to correct as they require either simultaneous estimation in multiple directions across the FoV \citep{Tasse2017}, or an iterative approach like ``peeling'' \citep{Noordam2004, vanWeeren2016b}. For this paper, we observed only fields whose total flux density is strongly dominated by a central compact source. As a consequence, also solutions are dominated by the systematic effects from \textit{that} source direction. Therefore, DDE can be treated as direction independent errors (DIE) whose values are an approximation of the DDE value in the direction of the dominant source.

\begin{figure*}[ht!]
  \centering
  \includegraphics[width=.9\textwidth]{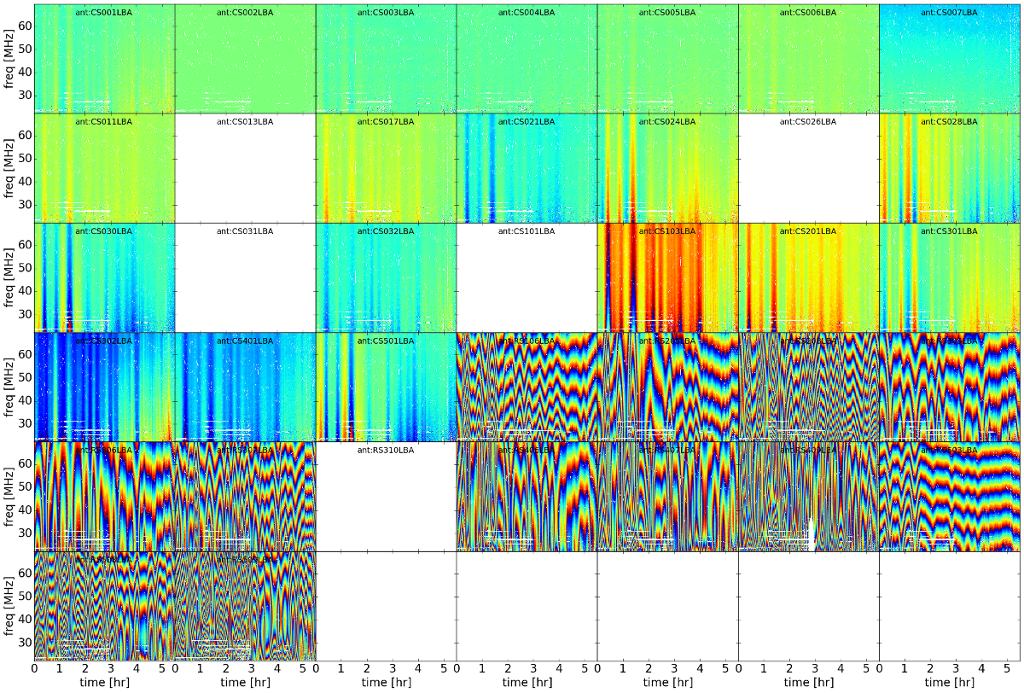}\\
  \includegraphics[width=.9\textwidth]{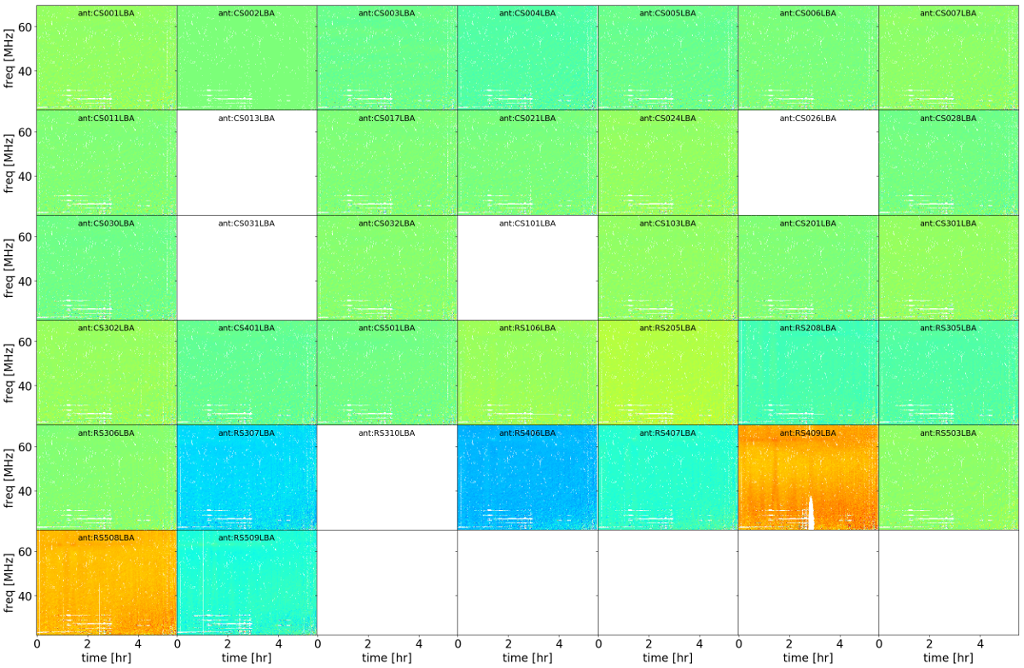}
  \caption{Gain phase solutions (from $+\pi$: blue to $-\pi$: red) for the XX polarisation referenced to station CS\,002 (located in the superterp centre). Each panel is one station. White pixels represents bad data that were removed. Stations CS\,013, CS\,026, CS\,031, CS\,101 and RS\,310 were fully removed due to hardware issues or strong radio frequency interference (RFI) contamination. In the bottom image we show the residuals after subtracting all the effects shown in Fig.~\ref{fig:effects}.}
  \label{fig:ph}
\end{figure*}

\begin{figure*}[ht!]
  \centering
  \includegraphics[width=.49\textwidth]{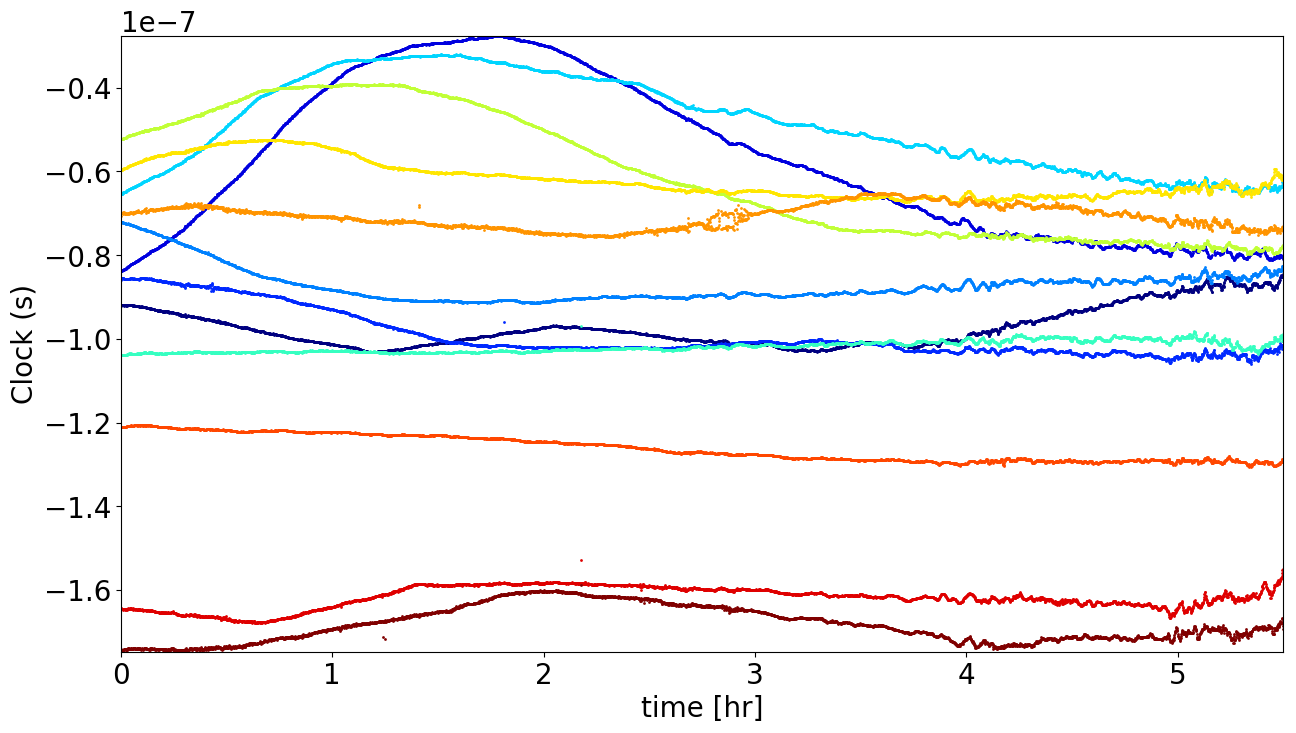}
  \includegraphics[width=.49\textwidth]{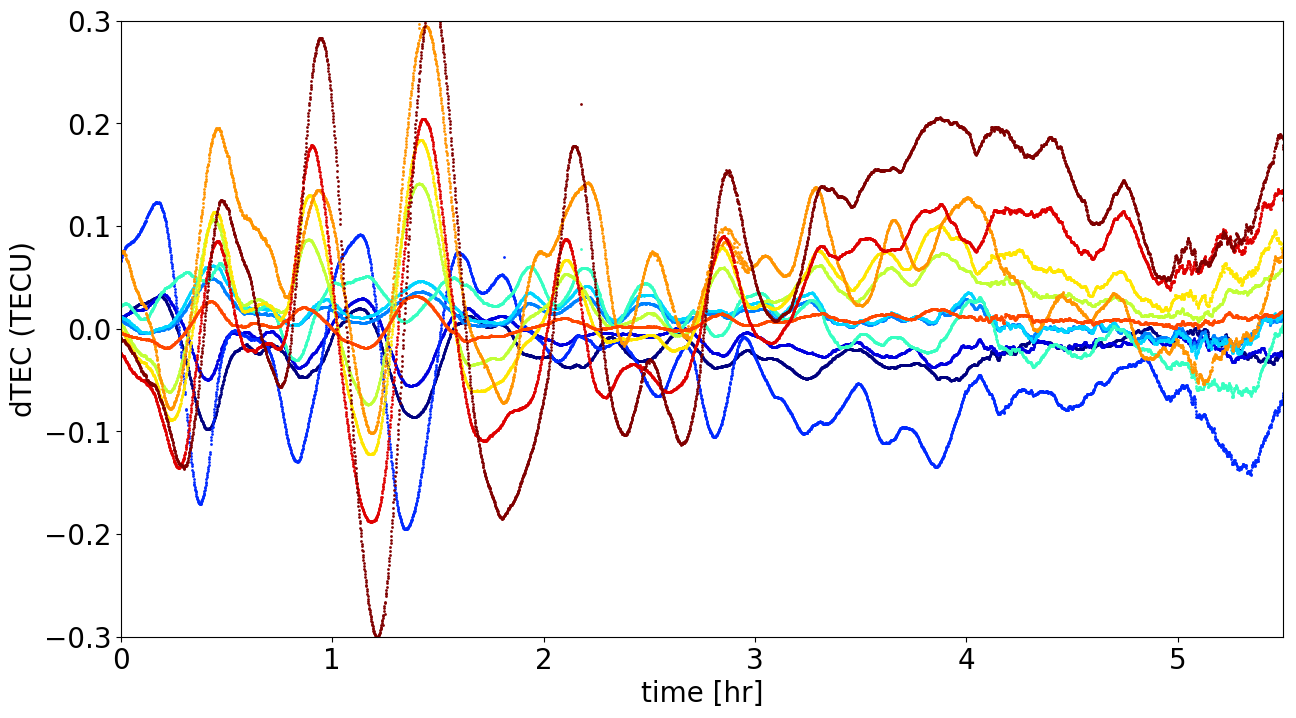}\\
  \includegraphics[width=.49\textwidth]{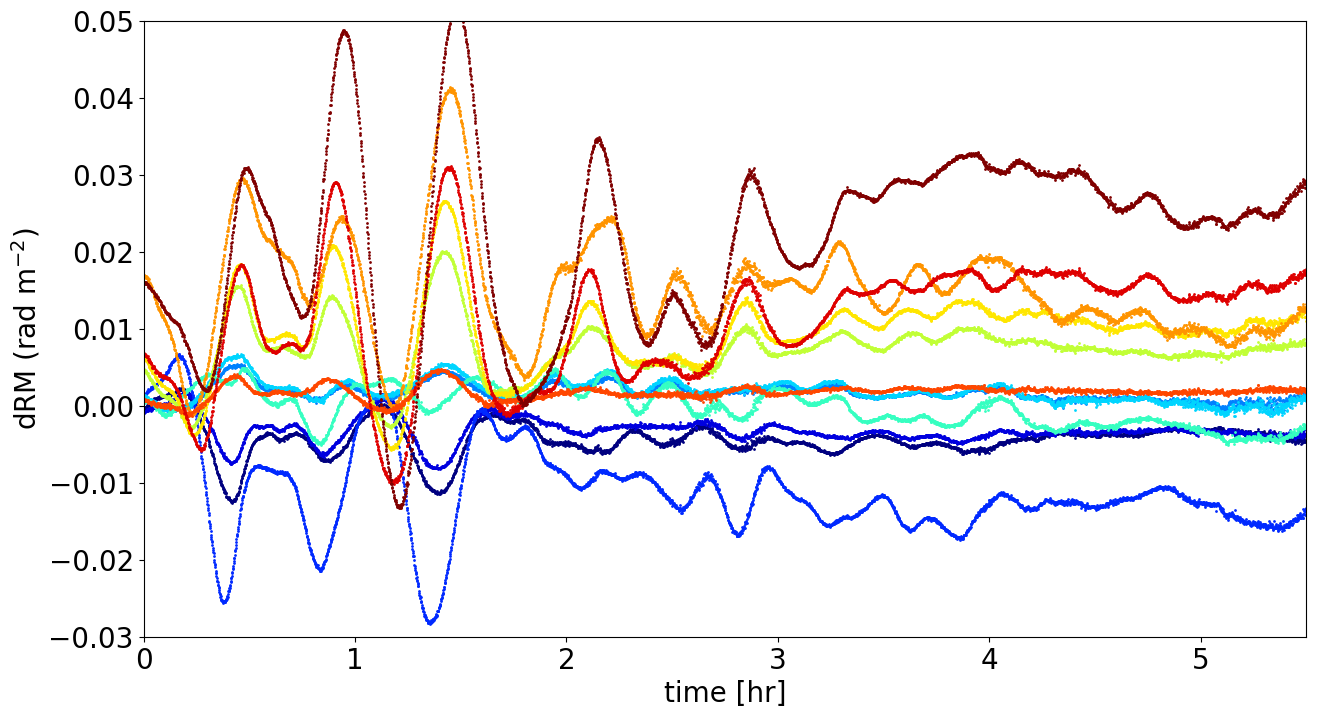}
  \includegraphics[width=.49\textwidth]{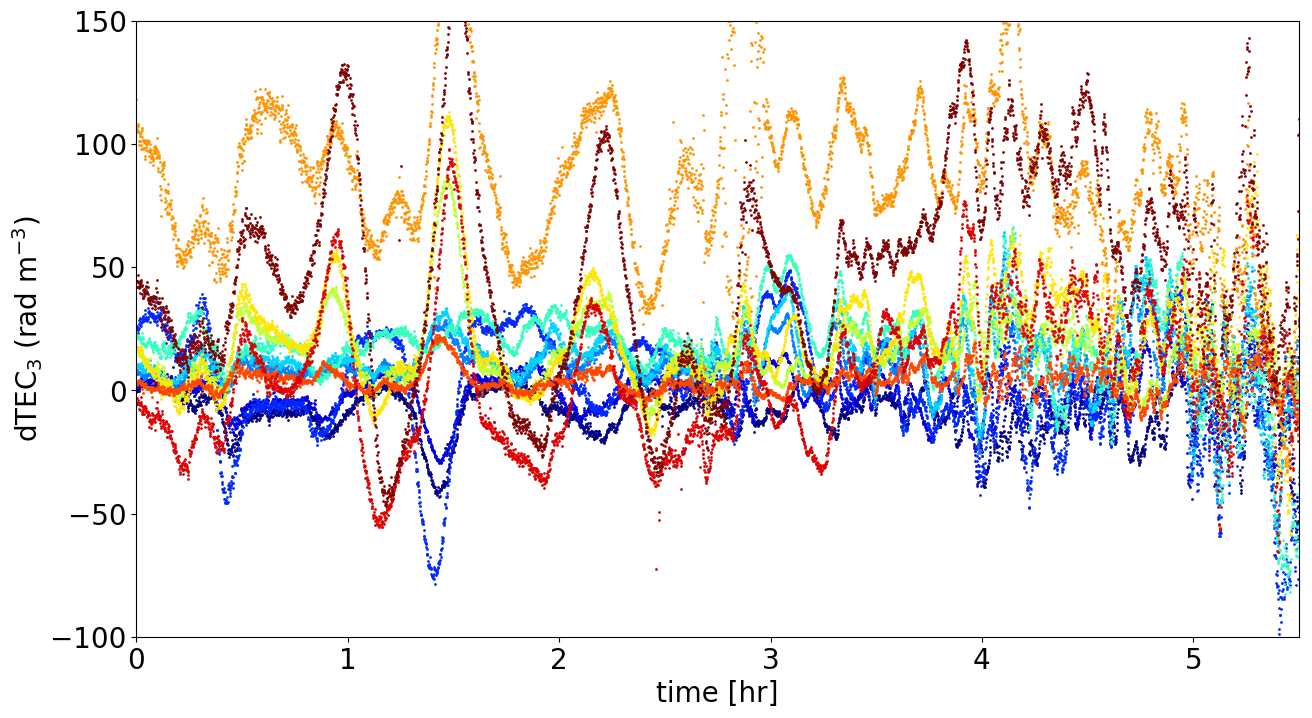}
  \caption{From top-left to bottom-right: instrumental clock delay (in s). Total electron content variation along the observation (in TECU). Faraday rotation (in rad m$^{-2}$). Ionospheric 3rd order effect (in rad m$^{-3}$). All values are differential between CS\,001 (assumed constant at 0) and all remote stations (from blue to red in alphabetical order).}
  \label{fig:effects}
\end{figure*}

\section{Systematic effects}

In Fig.~\ref{fig:ph} we show phase solutions for each station for 5.5 hrs of observation of 3C196, in the frequency range $22-70$ MHz. All stations labelled CS are ``core stations'' --- these stations share the same clock and are close together. Stations labeled RS are ``remote stations'' and each of them has an independent clock. Clocks are not perfectly synchronised and they might drift in time with respect to each other. This imprints a time-variable systematic error on the phases of remote stations which is linear in frequency ($\propto \nu$). This effect dominates the phase error in the closest remote stations such as RS\,106 that in fact shows a rather uniform phase wrapping across the sampled bandwidth. In the most distant remote stations (such as RS\,508 or RS\,509) the ionospheric dispersive delay ($\propto 1/\nu$) dominates the error budget, and as a consequence phases wrap differently at different frequencies. We can use the different frequency dependency of the two effects (clock errors and ionospheric dispersive delay) to disentangle their contribution to the phase solutions. This process is known as clock/TEC separation and is performed by the LOFAR Solution Tool (LoSoTo\footnote{\url{https://github.com/revoltek/losoto}}). The outcome is time-streams of values representing the various systematic effects that we want to disentangle. It is important to note that the clock/TEC separation procedure does not impose any time coherency in the expected systematic effects; each time stamp is treated separately. The fact that the outcome looks well-correlated in time is an indicator that the procedure works as expected. As a final verification we also subtracted the derived phase effect of all the systematic effects combined from the original solutions, this produced the second panel of Fig.~\ref{fig:effects} which shows rather uniform residuals, meaning that the vast majority of the LOFAR phase systematic effect can be described with a simple model of clock delays and ionospheric effects.

\subsection{Ionospheric (dispersive) and instrumental delays}

In the top two panels of Fig.~\ref{fig:effects} the clock delay and the ionospheric delay are plotted for each station after separation. These values are differential with respect to CS\,002, and as a consequence core stations (not plotted) have very small dTEC and a constant differential ``clock'' due to other instrumental delays. Conversely, remote stations show dTEC values larger than 0.3 TECU for stations that are around 50 km away from CS\,002. Furthermore, a strong correlation between the dTEC as measured by neighbouring stations is also visible.

The initial part of the observation is affected by large ionospheric traveling waves. We cross-correlated the dTEC values for RS\,407, RS\,508, and RS\,509 to estimate the wave direction and speed. We obtained a time lag of 137 s (RS\,509--RS\,508), 177 s (RS\,509--RS\,407), and 47 s (RS\,508--RS\,407). These values are compatible with waves traveling at $\sim 200$ m s$^{-1}$ from south-west to north-east, values compatible with previous measurements \citep[e.g.][]{Fallows2016}. The speed of these waves might be compatible with the propagation of traveling ionospheric disturbances (TIDs) even if the direction of the propagation is usually equatorward \citep{Cesaroni2017}. In fact, TIDs are associated with traveling atmospheric disturbances (TAD) originated at auroral latitude by Joule heating \citep{Hunsucker1982}. TIDs can have periods of minutes to hours \citep{Hocke1996}. In the second half of the observation the ionosphere becomes less ordered and likely more dominated by turbulent motions. Large variations in the dTEC are not necessarily an indication of a dataset that is difficult to calibrate. An ionosphere that is very active, but rather coherent across the FoV and with relatively slow variation in time (the first half of the test dataset) is easier to calibrate than an ionosphere that is highly direction dependent with fast scintillations (the second half of the test dataset).

\begin{figure*}[t]
  \centering 
  \includegraphics[width=.49\textwidth]{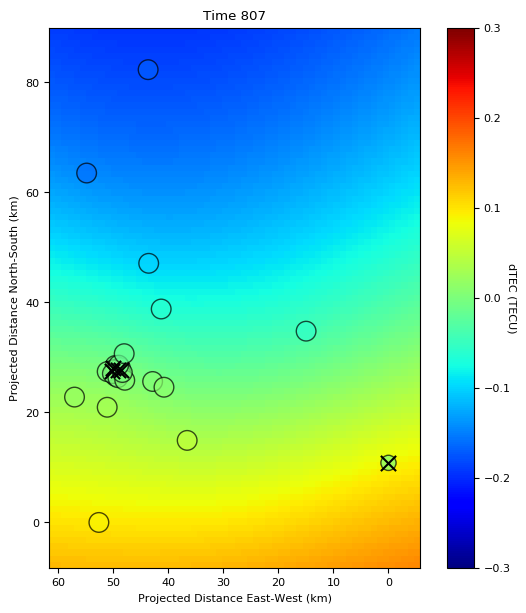}
  \includegraphics[width=.49\textwidth]{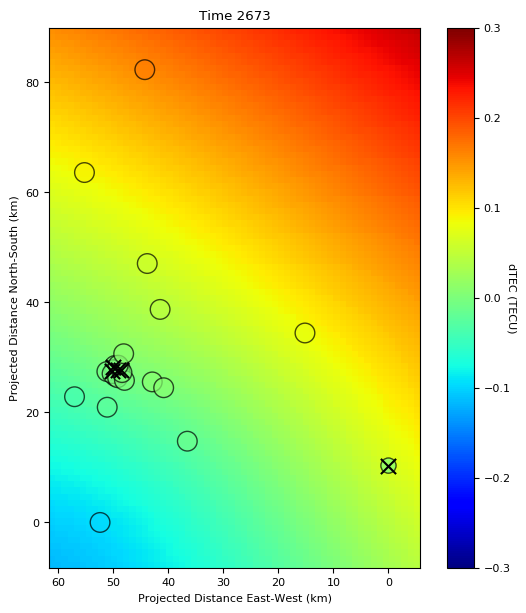}
  \caption{Two examples of TEC-screen fits from the observation shown in Fig.~\ref{fig:effects}. Values are in differential TECU with respect to station CS\,002. Each circle represents a LOFAR station; colour coded within the circle is the dTEC value for that station (note that the markers are \textit{filled} with the colour corresponding to the measured dTEC). The background colour is the TEC screen fitted across the array as described in the text. Stations crossed with an X are the same excluded from Fig.~\ref{fig:ph}. The full movie is available in the online material.}
  \label{fig:screen}
\end{figure*}

To demonstrate the spatial coherence of the dTEC solutions, we construct spatial screens of the dTEC variations across the array. One screen is made for each time slot solved for in the calibration, fitting to the projected locations (pierce points) of 3C196 onto a thin plane located at a height of 200 km above the ground. Generally, there is one pierce point per station, per direction, per time slot. In the case of the fits discussed here, we use a single direction (that of 3C196); therefore we can have up to 38 pierce points per screen (some of which are flagged; see Fig.~\ref{fig:ph}). We adopt the same method of \citet{Intema2009} that uses Karhunen-Lo\`{e}ve (KL) base functions to model the spatial variations of the dTEC values. During the fitting, we adopt a power-law dependence for the phase structure function (see Section~\ref{sec:structfunc}), with a power-law index, $\beta = 5/3$. Lastly, we fix the number of KL base vectors fit per screen to 5, resulting in 5 free model parameters per screen.

We plot the screen fits at two times in Fig.~\ref{fig:screen}. The spatial coherence of the dTEC values at each time is clearly visible in these plots. Therefore, it should be possible to interpolate spatially between pierce points using the dTEC screens to obtain dTEC values in directions other than that of 3C196. The screens can also be used during fitting as constraints that enforce spatial coherence of the dTEC solutions, thus reducing the number of free parameters solved for during the calibration. We are currently investigating the use of screens in this way.

\subsection{Higher order terms}

\begin{figure*}[ht]
  \centering
  \includegraphics[width=\textwidth]{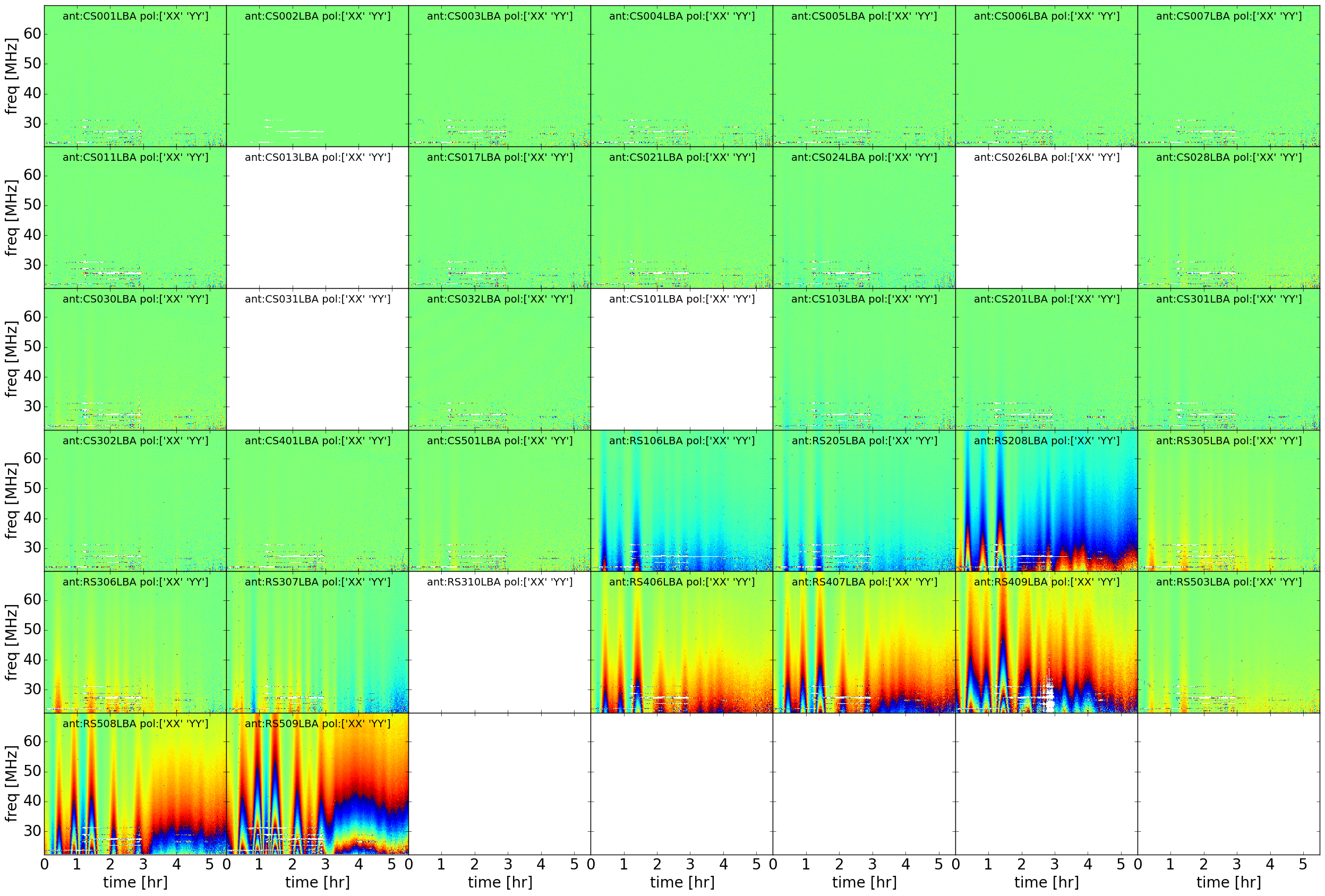}
  \caption{Same as in Fig.~\ref{fig:ph}, but here we show the differential phase solutions in circular polarisation (RR -- LL). The plot isolates the effect of Faraday rotation. Since it depends on dTEC and d$B$, the effect is stronger on stations further away from the reference (CS\,002). By fitting a $1/\nu^2$ dispersive delay we can recover the rotation measure displayed in the third panel of Fig.~\ref{fig:effects}.}
  \label{fig:phrot}
\end{figure*}

With this dataset we were able to measure the second order ionospheric effect due to Faraday rotation and the third order dispersive delay effect. Faraday rotation is not easily obtainable from the XX and YY phase solutions, instead we took advantage of its different sign in the right and left polarisations (see Eq.~\ref{eq:n}). This required a conversion of the dataset from linear to circular polarisation and the extraction of the RR-LL phase solutions (see Fig.~\ref{fig:phrot}). Since clock delay and ionospheric first and third terms are scalar, their effect is the same for the RR and LL polarisation and they cancel out if we subtract one from the other. At this point, one can now easily fit the second order term as a differential delay between the two polarisations with a $1/\nu^2$ dependency. In the third panel of Fig.~\ref{fig:effects} we show the result in terms of rotation measure ($\rm RM$).

\begin{equation}
 \Phi = {\rm RM}\,\lambda^2 = \frac{2 \pi \nu}{c} \cdot \left( \frac{q^3}{16\pi^3 m_e^2 \epsilon_0} \cdot \frac{1}{\nu^3} \int_{\rm LoS} n_e B \cos\theta\,{\rm d}l \right).
\end{equation}

The term in parentheses comes from the second term of Eq.~\ref{eq:n}. Again we observe that nearby stations have a similar behaviour and that there is a correlation between the dTEC and the dRM. This is a consequence of the presence of $n_e$ in both the first and second terms of Eq.~\ref{eq:n}. The measured RM of 0.03 rad produces $\sim45\deg$ phase error, which is compatible with expectations given a measured dTEC of 0.3 TECU.

In certain observations, such as the one used in this example, when the ionospheric variation is strong enough, the third order term can also be extracted by fitting a $1/\nu^3$ term together with the clock delay and the ionospheric first-order dispersive delay at the clock/TEC separation time. The third order term becomes relevant only below $\sim40$~MHz but it can rarely be ignored at frequencies close to the plasma frequency ($<20$~MHz). In the last panel of Fig.~\ref{fig:effects} we show the estimated third order effect using:

\begin{equation}
 \Phi = {\rm TEC_3}\,\lambda^3 \approx \frac{2 \pi \nu}{c} \cdot \left( \frac{q^4}{128 \pi^4 m_e^2 \epsilon_0^2} \cdot \frac{1}{\nu^4} \int_{\rm LoS} n_e^2 \,{\rm d}l  \right).
\end{equation}

As the first and third order terms both depend on the number of free electrons $n_e$, not surprisingly they trace each other. In the initial part of the observation the effect is stronger due to the large dTEC and possibly other ionospheric characteristics (e.g. a thin dominant layer). In the second part of the observation the third order term becomes less prominent and harder to fit. As a consequence imperfect separation in the parameter determination generates small noise-like perturbations in the other estimated parameters. The second order term is not affected by this problem as its estimation is done with different methods as described above.

\subsection{Amplitude scintillations}

\begin{figure*}[t]
  \centering
  \includegraphics[width=.195\textwidth]{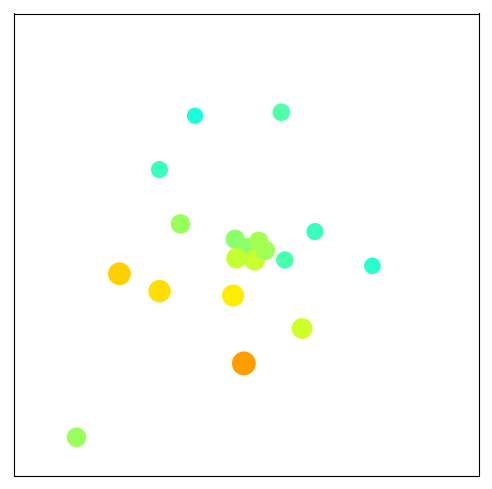}
  \includegraphics[width=.195\textwidth]{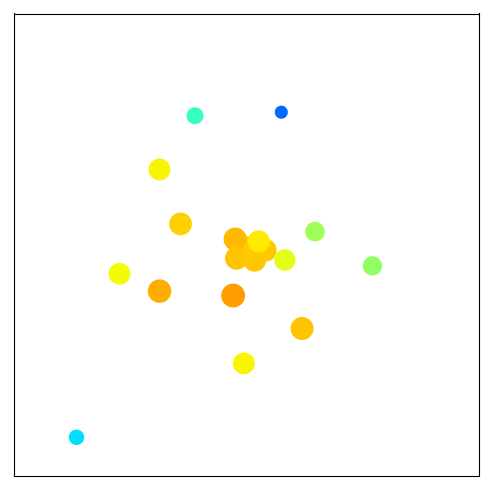}
  \includegraphics[width=.195\textwidth]{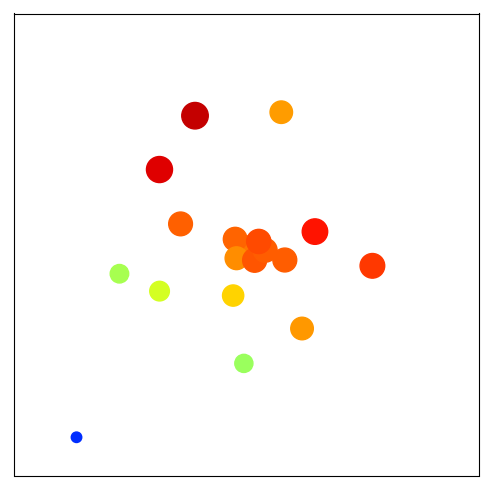}
  \includegraphics[width=.195\textwidth]{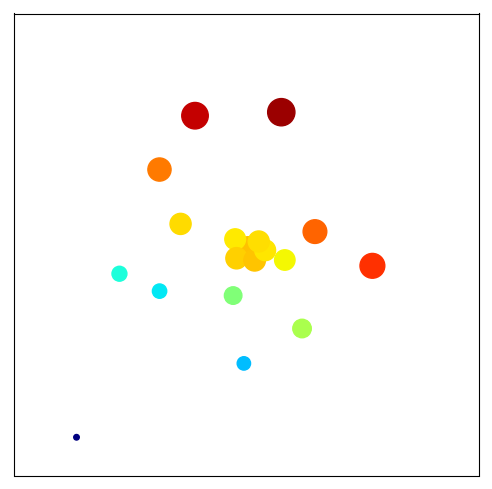}
  \includegraphics[width=.195\textwidth]{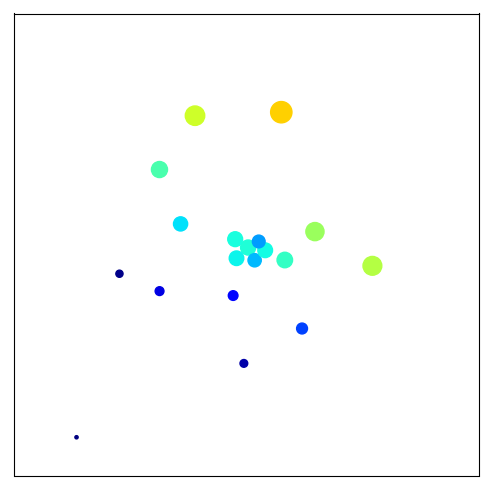}
  \put(-515,90){T1}\put(-410,90){T2}\put(-305,90){T3}\put(-200,90){T4}\put(-95,90){T5}
  \caption{Left to right: time evolution of amplitude solutions for LOFAR core stations for the observation in shown in Fig.~\ref{fig:effects}. Each plot is separated from the next by 10 seconds. Each point represents a core station and it is positioned in the plot according to its geographical location. Colour (blue to red) and size (small to large) are related to the value of the amplitude correction (from $-10\%$ to $+10\%$). A wave travelling south-north is visible, similar structures were also reported with single station observations (priv. comm. Richard Fallows). The full movie is available in the online material.}
  \label{fig:scint}
\end{figure*}

Scintillations are caused by electromagnetic waves scattered in a non uniform medium with small changes in the refractive index, such as the ionosphere. A plane wave that enters such a medium with a spatially uniform phase, exits the medium with a spatially irregular phase. After propagation to a station, the irregular phases may combine either constructively or destructively \citep[see][]{Kintner2007}. As a consequence, the wave amplitude is increased or decreased and the gain amplitude solutions of the station compensate for the effect by producing an exact opposite trend. To some degree, at frequencies below 100 MHz, we observe that scintillations are always present in LOFAR amplitude solutions.

An example of this is visible in Fig.~\ref{fig:scint}, where an ``amplitude wave'' crosses the core area (4 km) in 1 minute, therefore travelling at a velocity of $\sim240$~km/h. Projecting the linear size of these coherent structures to the height of 300 km, we can estimate an angular size over which amplitude corrections can be considered fairly uniform, which is $\sim 20$\arcmin. Scintillations therefore create a strongly direction-dependent amplitude error across the FoV.
%
%

Interestingly, wave-like structures are a recurrent pattern across the core stations in the second half of our test observation (see the online material). On the other hand, in the first part of the observation that is dominated by large waves, amplitude scintillations seem to be quasi-simultaneous across the entire array (including remote stations). The amplitude of the scintillations is around $\pm10\%$ during both halves of the observation; therefore, it does not seem to depend strongly on the size of the wave phenomena. Although scintillations with timescales larger than a few seconds are highly unlikely to be due to the interplanetary medium, the possibility cannot be entirely ruled out: \citet{Kaplan2015} demonstrated that interplanetary scintillation (IPS) was still visible in observations taken with the Murchison Widefield Array (MWA, \citet{Tingay2012}) with a time resolution of 2s, and \citet{Fallows2016} demonstrated that IPS appears near-simultaneous across the Dutch stations of LOFAR.  Furthermore, dedicated observations of IPS taken with LOFAR show good IPS signal for the radio sources observed here, with both 3C295 and 3C380 used in an IPS observing campaign run during October 2016 and 3C196 used in spring and summertime observations.

\section{The ionosphere as sensed by LOFAR}
\subsection{Structure function}\label{sec:structfunc}

\begin{figure}[t]
  \centering
  \includegraphics[width=.5\textwidth]{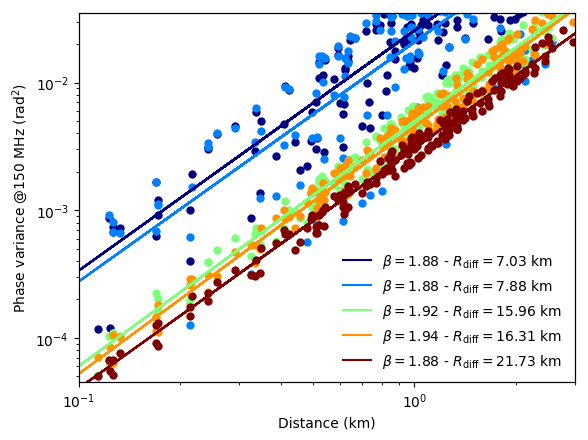}
  \caption{Phase structure function divided into five time chunks for the observation of Fig.~\ref{fig:effects}. The initial part of the observations is visibly more affected by ionospheric disturbances also in Fig.~\ref{fig:effects}. The phase variance is converted to the expected value at 150~MHz to compare it with other experiments such as LOFAR HBA or MWA.}
  \label{fig:sfunct}
\end{figure}

We divided the observations into five time chunks of $\sim 1$ hr each and calculated the ionosphere phase structure function as described in \cite{Mevius2016}. For Kolmogorov turbulence, the phase structure function is a power-law of the form:

\begin{equation}\label{eq:diffsc}
 D(r) = \left(\frac{r}{r_{\rm diff}}\right)^\beta.
\end{equation}

We fit this function to the data to obtain an estimation of $\beta$ (expected to be $5/3=1.67$ for pure Kolmogorov turbulence) and of $r_{\rm diff}$, the spatial scale over which the phase variance is 1 rad$^2$, and is referred to as the diffractive scale \citep{Narayan1992}. Since the phase errors of the core stations are dominated by ionospheric errors, we did not use those phases directly, instead first converting to  dTEC. This gives a fast rough estimate of the diffractive scales and thus of the ionospheric quality during an observation. The air mass factor was not corrected for, resulting in a slightly larger variance depending on the elevation angle. We obtained $\beta = 1.88,1.88,1.92,1.94,1.88$ and $r_{\rm diff@150 MHz} = 7,8,16,16,22$~km, for time chunks 1 to 5, respectively (see Fig.~\ref{fig:sfunct}). These values for $\beta$ are larger than expected for Kolmogorov turbulence ($\beta =5/3$), and are probably due to the effect of large scale waves, which were not filtered. The baselines that were used for Fig.~\ref{fig:sfunct} are much smaller than the typical wavelength of such waves ($\sim 100-200$ km), therefore the effect closely resembles that of a linear gradient over these baselines, which would correspond to $\beta=2.0$ in the phase structure function. The results are similar to those reported in \cite{Mevius2016}. The smaller diffractive scales correspond to the first half of the observation, where the magnitude of the dTEC variations is larger. 

After removing the large ionospheric gradient (e.g. with a direction independent calibration), higher orders of the refractive index expansion in Eq.~\ref{eq:n} are small, and are assumed to be uniform across the LOFAR beam. To test this assumption we can use the structure function. The LOFAR primary beam size is $\sim4\deg$, that corresponds to 20 km at a 300 km high ionospheric layer. With a diffractive scale of 10 km and assuming $\beta=1.7$, Eq.~\ref{eq:diffsc} gives a phase variance of $\sim3$ rad$^2$ at 150 MHz. This corresponds to a dTEC of about 0.03 TECU. Consequently, even at 60 MHz, only the first order term varies substantially across the FoV while other terms can be considered constant. This finding can be used to simplify the calibration strategies on the target fields.

\subsection{Multi-epoch observations}

\begin{landscape}
\begin{figure}
  \centering
  \includegraphics[width=1.2\textheight]{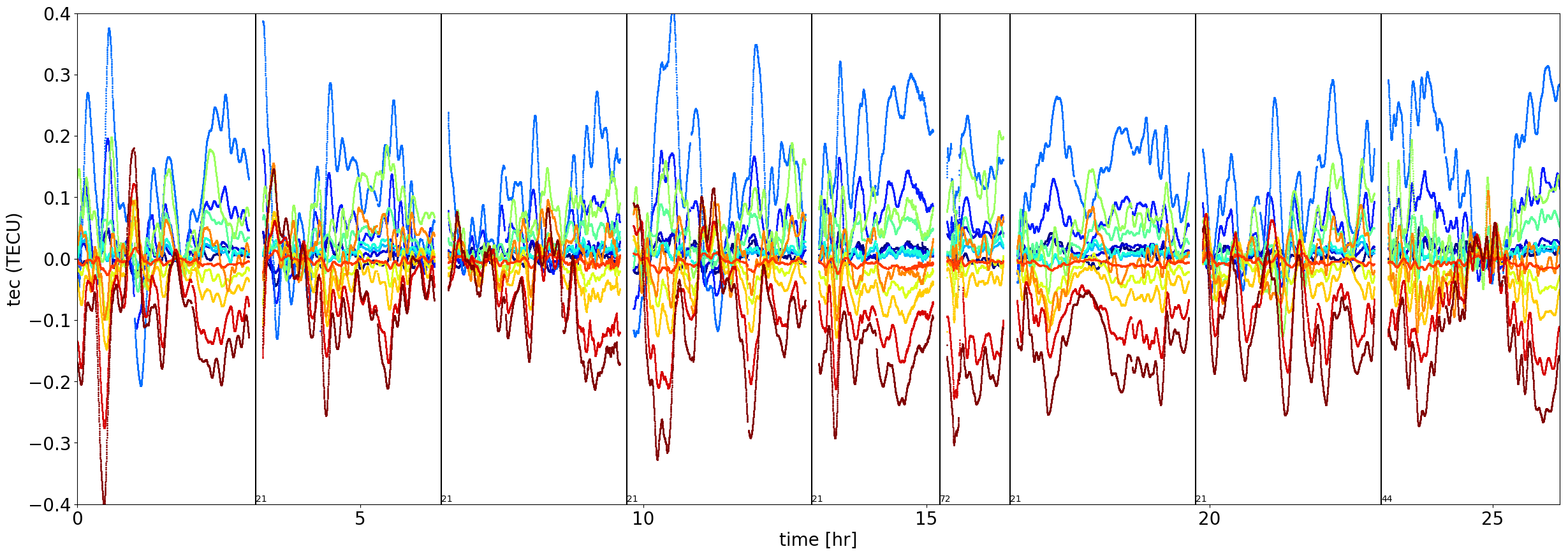}\\
  \includegraphics[width=1.2\textheight]{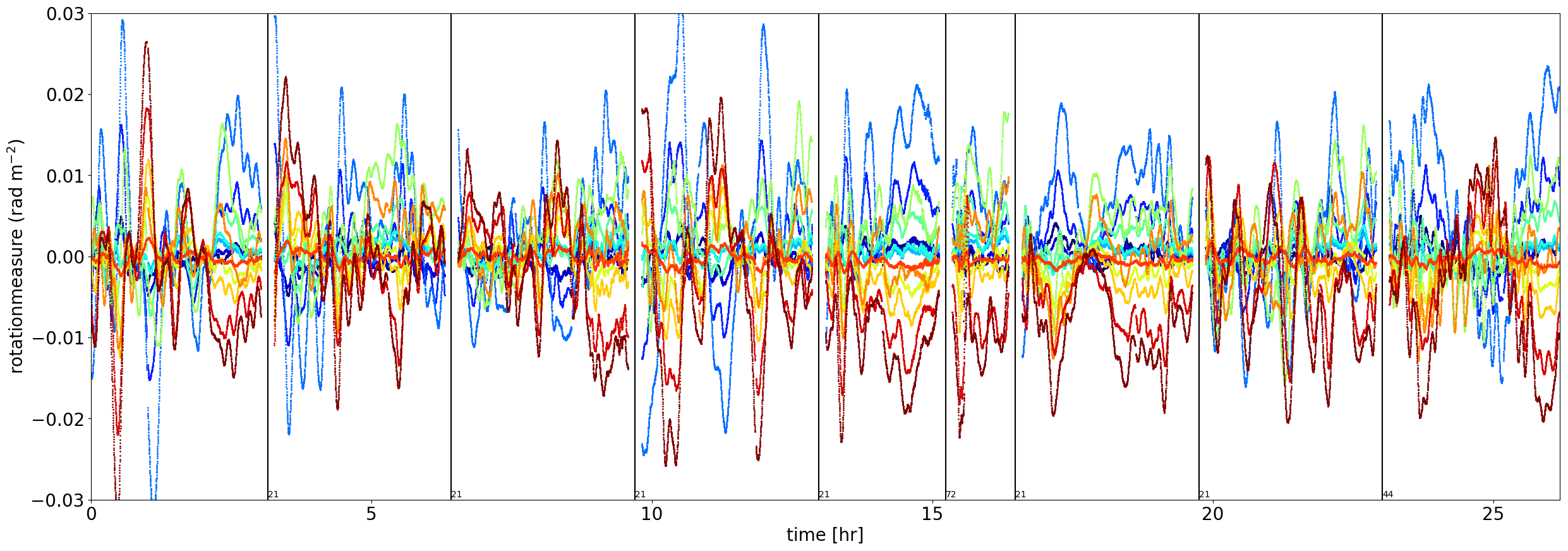}
  \caption{First and second order ionospheric effects quantified by variation of differential TEC and differential Faraday rotation between the LOFAR core and all remote stations. Each panel is a separate observation towards 3C380. The gaps between observations are: 21, 21, 21, 21, 72, 21, 21, and 44 hrs. Stations are colour-coded in alphabetical order. The similarity between the top and bottom panels is due to the TEC dependency of Faraday rotation.}
  \label{fig:3c380}
\end{figure}
\end{landscape}

In November 2017 we performed a series of 8 observations pointed at the calibrators 3C295 and 3C380. The observations were carried out during day-time. We extracted the dTEC and dFR measurements for all of them (see Fig.~\ref{fig:3c380} and Fig.~\ref{fig:3c295}). A basic expectation of ionospheric models is the presence of an increasing north-south TEC gradient. In all observations the dTEC values shows its presence (red lines are stations in the North, blue lines are stations in the South, all referenced to station CS\,002 in the superterp). The wave-like behaviour of the ionospheric systematic effects are clear although during some time intervals waves appear more structured and on larger scales than at other times \citep[as seen with different techniques with MWA][]{Loi2015a,Loi2015b}.

Ionospheric conditions are driven by a number of factors (e.g. season, solar activity, and latitude). The dominant factor is the day-night cycle. During night-time the ionospheric TEC is reduced; this has an impact on the magnitude of the second order term of Eq.~\ref{eq:n}, which is expected to be smaller.  However, single-station observations taken under an ionospheric scintillation monitoring project demonstrate that scintillation is often much stronger at night, particularly prior to midnight.  We report a large fraction of data loss due to intense scintillation events during night-time observations.  During these events we were not able to disentangle the various phase effects, and amplitude solutions show clear signs of decorrelation. Along a 7-night campaign we did in February and March 2015 around 30 per cent of our observations had to be discarded for this reason. Conversely, in 14 days of day-time observations taken in 2017 (seven of which are presented in this paper) only a few per cent of our data were affected by strong scintillations. Since the observations were taken far apart in time, it is difficult to derive strong conclusions. A downside of day-time observations is the (unavoidable) presence of the Sun, combined with the relatively poor side-lobe suppression of a phased array like LOFAR compared to dish-based radio interferometers. However, the low-frequency radio spectrum of the (quiet) Sun is strongly inverted, with $S_\nu \propto \nu^{3}$, and as a consequence at 50 MHz the Sun flux density is expected to be a few thousand Jy, which is comparable to the flux density of some other sources present at night-time. Moreover, solar emission is extended on scales of several arcmin; therefore, it is resolved out on moderately long baselines.

\begin{figure}[t]
  \centering
  \includegraphics[width=.5\textwidth]{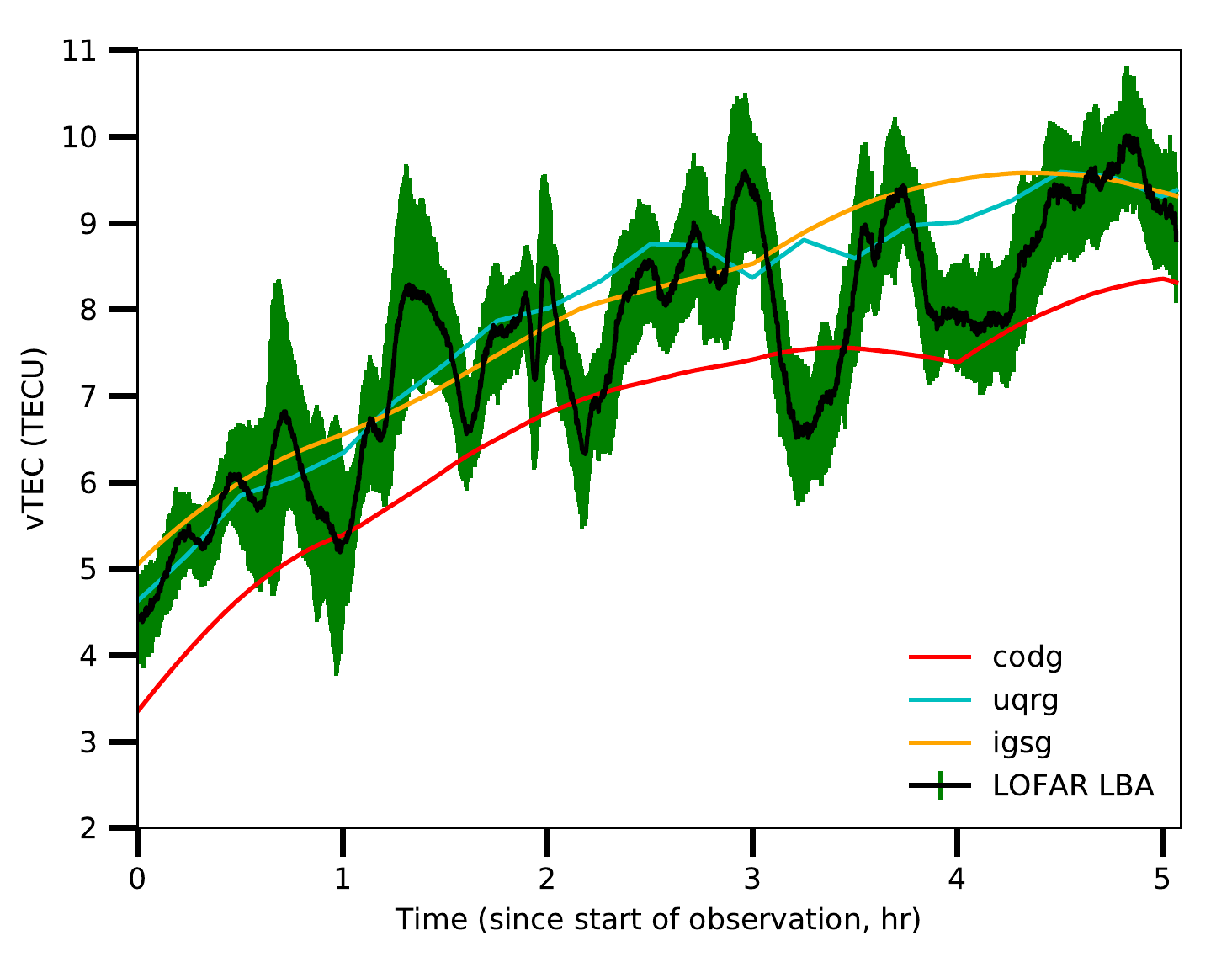}
  \caption{Absolute TEC derived from World Magnetic Model, dTEC and dRM of the first observation in figure \ref{fig:3c295}. Compared to three different interpolated IGS model datasets.}
  \label{fig:absoluteTEC}
\end{figure}

\subsection{Absolute TEC and satellite comparison}

\begin{landscape}
\begin{figure}
  \centering
  \includegraphics[width=1.2\textheight]{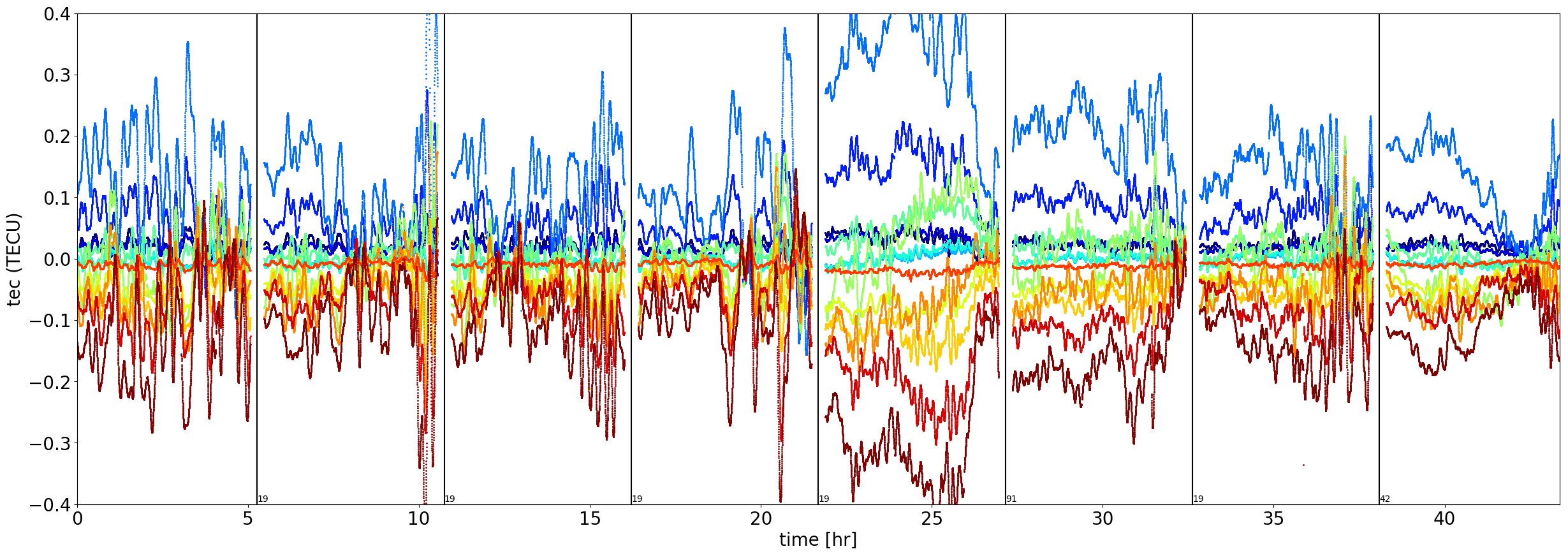}\\
  \includegraphics[width=1.2\textheight]{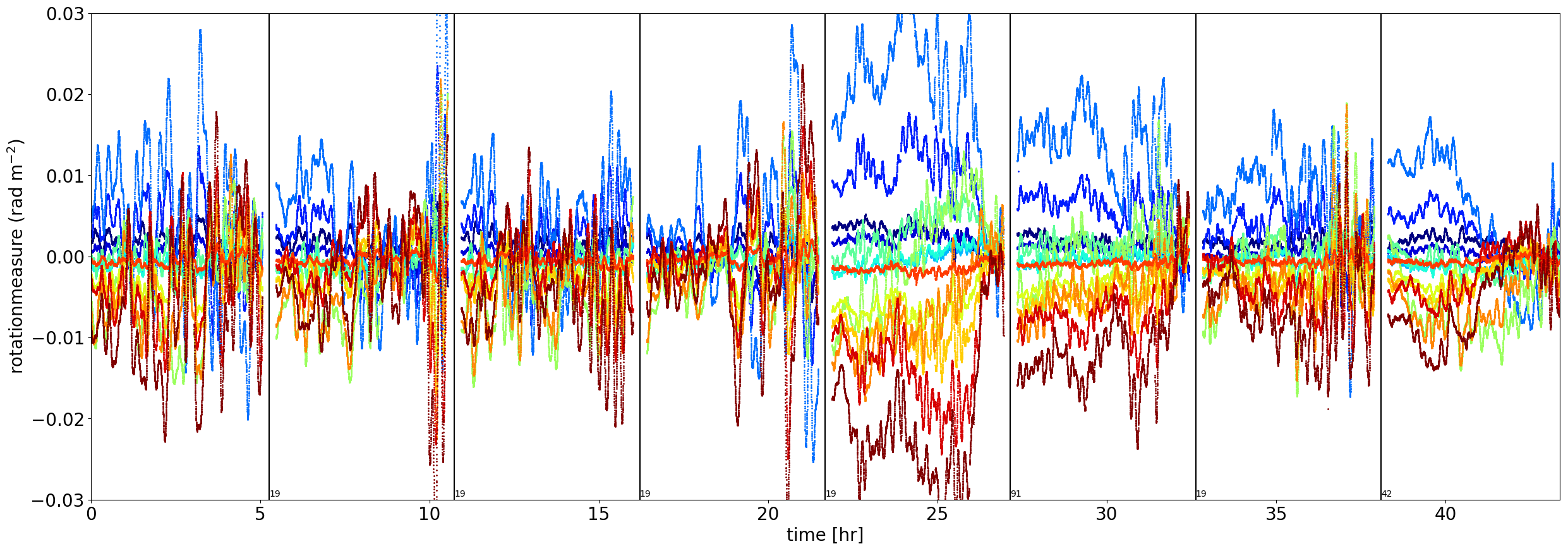}
  \caption{Same as Fig.~\ref{fig:3c380} but for 3C295. Gaps between observations are: 19, 19, 19, 19, 91, 19, and 42 hrs.}
  \label{fig:3c295}
\end{figure}
\end{landscape}

An interesting application is the
possibility to estimate the absolute TEC from the differential TEC and
differential Faraday rotation. We make the following approximation:
\begin{equation}
{\rm dRM}_{12} \approx c\cdot B||_1\cdot {\rm TEC}_1 -c\cdot B||_2 \cdot {\rm TEC}_2,
\end{equation}
with ${\rm dRM}_{12}$ differential Faraday rotation between station 1 and 2  and
$B||_{1,2}$ the parallel magnetic field and ${\rm TEC}_{1,2}$ the integrated
electron content along the LoS for station 1 and 2 respectively. We approximate the integral in equation \ref{eq:delay} with the thin layer approach. This can be rewritten in terms of absolute TEC at station 1:
\begin{equation}\label{eq:absoluteTEC}
{\rm TEC}_1\cdot {\rm d}B||_{12} = C\cdot {\rm dRM}_{12} - B||_2\cdot {\rm dTEC}_{12},
\end{equation}
with ${\rm d}B||_{12}$ the differential parallel magnetic field and ${\rm dTEC}_{12}$ the
differential integrated electron content between station 1 and 2. We used the
World Magnetic Model \citep{Chulliat2014} and the measurements of dRM and dTEC to
estimate the absolute vertical TEC for the first observation in figure \ref{fig:3c295} in this way. The thin layer model places a single ionospheric layer at an altitude of 450 km. Slant
TEC to vertical TEC conversions are done assuming a spherical ionosphere at the
same altitude. The resulting vertical TEC at the position of CS\,001 as a
function of time is  shown in Fig.~\ref{fig:absoluteTEC}. The error bars
reflect the spread of the measurements using all different station
combinations. In the same plot the interpolated vertical TEC values from three different Global TEC maps provided by International GNSS service (IGS) are shown. In particular, CODE, UPC and combined final products have been used \citep{Schaer1999,Orus2005,Hernandez-Pajares2009}. The $B$ field parameters and satellite based TEC values were determined using the \texttt{RMextract} package\footnote{\url{https://github.com/lofar-astron/RMextract}.}. The measured absolute TEC values are within $10\%$ and following the trend of the GNSS based products from IGS. The time
resolution is 10 s, 2 orders of magnitude better than that of the IGS models which is 2 hrs for codg  and igsg and 15 minutes for uqsg.  The main uncertainty in the LOFAR measurement comes from the value of ${\rm d}B||$, which can be very small. A small unmodelled perturbation of ${\rm d}B||$ could artificially enhance the amplitude of the waves as observed in figure \ref{fig:absoluteTEC}. Here we merely illustrate the principle of concept. A further discussion of systematic uncertainties of this method will be the subject of a subsequent publication (Mevius et al. in preparation).

\section{Conclusions}

The ionosphere is the main limiting factor of the quality of low-frequency radio-interferometric observations. The time/space variable refractive index of the ionospheric plasma generates highly direction-dependent dispersive delays that affect phases recorded by the interferometer. At ultra-low frequencies differential Faraday rotation and higher-order terms (in frequency) also become prominent. The most important results of this paper are:
\begin{enumerate}
 \item We showed that LOFAR station-based gain phase can be decomposed into a small number of systematic effects: clock delays, ionospheric effects of 1st, 2nd (Faraday rotation), and 3rd order. The 3rd order effect is important only for observations below $\sim40$ MHz.
 \item We showed that the ionospheric parameters we derived with our decomposition are consistent with expectation (e.g. they are time and spatially coherent) and with independent measurements from satellites. This procedure also demonstrates that LOFAR can be used to obtain independent measurements of the absolute TEC.
 \item We showed that visibility amplitudes are affected by scintillations and that at ultra-low frequencies ($<100$ MHz) they are always present. We show that amplitude scintillations also follow patterns both in time and space and in certain periods behave like travelling waves across the array.
 \item We showed that scintillations at night-time can result in a data loss of $\sim 30\%$ and that this phenomenon is reduced during the day. Based on these results we would recommend to observe at daytime. However, the origin of the scintillations (interplanetary or ionospheric) needs to be clarified and more statistics need to be gathered for a definitive answer.
\end{enumerate}

We also gather here some consideration for future low-frequency experiments:
\begin{enumerate}
 \item Low-frequency telescopes operate in a regime of low signal-to-noise ratio due to the high sky temperature. The ionosphere itself contributes with one or more parameters to be estimated every few seconds. It is important that no further degrees of freedom are added to the solving process (e.g. by using different clocks in different stations).
 \item Again to maximise the signal-to-noise ratio modern solvers need to be conceived in order to exploit all known coherent structures of the ionosphere, including spatial, time and frequency coherency. In this paper time-coherency has never been imposed, although this has been already considered in some modern algorithms \citep[e.g.][]{Tasse2014}.
 \item Several regimes are identifiable based on the observing frequency and the maximum baseline length (see Tab.~\ref{tab:err}). The adopted calibration strategy must vary accordingly, considering, given the desired dynamic range, which effects are relevant and which have direction-dependent properties.
 \item The multi-beam capability of modern phased-array interferometers needs to be tested extensively to see how large a reliable TEC screen can be. Furthermore, this capability can be used to develop novel techniques to transfer time-variable instrumental errors from the calibrator to the target fields.
 \item So far, the use of GNSS data to calibrate radio interferometric observations has been marginally used. However, it would be important to check those data in real time to assess the ionospheric state and therefore decide whether to schedule ultra-low frequency observations (or any interferometric observation at all).
 \item In sight of the large data volume produced by SKA, an automatic pipeline to reduce calibrator data could assess the quality of the observation in a short time-period and provide a quantitative figure of merit to decide whether to archive the data or to re-schedule the observation (e.g. through plots similar to Fig.~\ref{fig:effects}).
\end{enumerate}

\begin{acknowledgements}

The authors want to thank Claudio Cesaroni for the many discussions and suggestions.

F.d.G. is supported by the VENI research programme with project number 1808, which is financed by the Netherlands Organisation for Scientific Research (NWO). M.M. acknowledges support from the ERC (grant 339743, LOFARCORE) .

LOFAR, the Low Frequency Array designed and constructed by ASTRON, has facilities in several countries, that are owned by various parties (each with their own funding sources), and that are collectively operated by the International LOFAR Telescope (ILT) foundation under a joint scientific policy. This work had made use of the Lofar Solution Tool (LoSoTo), developed by F. de Gasperin.

\end{acknowledgements}


\bibliographystyle{aa}
\bibliography{papers-iono}

\begin{thebibliography}{31}
\expandafter\ifx\csname natexlab\endcsname\relax\def\natexlab#1{#1}\fi

\bibitem[{Cesaroni {et~al.}(2017)Cesaroni, Alfonsi, Pezzopane, Martinis,
  Baumgardner, Wroten, Mendillo, Music{\`{o}}, Lazzarin, \&
  Umbriaco}]{Cesaroni2017}
Cesaroni, C., Alfonsi, L., Pezzopane, M., {et~al.} 2017, J. Geophys. Res. Sp.
  Phys., 122, 11,794

\bibitem[{Chulliat {et~al.}(2014)Chulliat, Macmillan, Alken, Beggan, Nair,
  Hamilton, Woods, Ridley, Maus, \& Thomson}]{Chulliat2014}
Chulliat, A., Macmillan, S., Alken, P., {et~al.} 2014, {The US/UK World
  Magnetic Model for 2015-2020, NOAA National Geophysical Data Center, Boulder,
  CO}

\bibitem[{Cohen {et~al.}(2007)Cohen, Lane, Cotton, Kassim, Lazio, Perley,
  Condon, \& Erickson}]{Cohen2007}
Cohen, A.~S., Lane, W.~M., Cotton, W.~D., {et~al.} 2007, Astron. J., 134, 1245

\bibitem[{Datta-Barua {et~al.}(2008)Datta-Barua, Walter, Blanch, \&
  Enge}]{Datta-Barua2008}
Datta-Barua, S., Walter, T., Blanch, J., \& Enge, P. 2008, Radio Sci., 43,
  RS5010

\bibitem[{Davies(1990)}]{Davies1990}
Davies, K. 1990, {Ionospheric Radio}, iee electr edn., Vol.~36 (Peter Pergrinus
  Ltd., London, UK)

\bibitem[{Fallows {et~al.}(2016)Fallows, Bisi, Forte, Ulich, Konovalenko, Mann,
  \& Vocks}]{Fallows2016}
Fallows, R.~A., Bisi, M.~M., Forte, B., {et~al.} 2016, Astrophys. J., 828, L7

\bibitem[{Hernandez-Pajares {et~al.}(2009)Hernandez-Pajares, Juan, Sanz, Orus,
  Garcia-Rigo, Feltens, Komjathy, Schaer, \&
  Krankowski}]{Hernandez-Pajares2009}
Hernandez-Pajares, M., Juan, J.~M., Sanz, J., {et~al.} 2009, J. Geod., 83, 263

\bibitem[{Hocke \& Schlegel(1996)}]{Hocke1996}
Hocke, K. \& Schlegel, K. 1996, Ann. Geophys., 14, 917

\bibitem[{Hoque \& Jakowski(2008)}]{Hoque2008}
Hoque, M.~M. \& Jakowski, N. 2008, Radio Sci., 43, n/a

\bibitem[{Hunsucker(1982)}]{Hunsucker1982}
Hunsucker, R.~D. 1982, {Atmospheric gravity waves generated in the
  high‐latitude ionosphere: A review}

\bibitem[{Hurley-Walker {et~al.}(2017)Hurley-Walker, Callingham, Hancock,
  Franzen, Hindson, Kapinska, Morgan, Offringa, Wayth, Wu, Zheng, Murphy, Bell,
  Dwarakanath, For, Gaensler, Johnston-Hollitt, Lenc, Procopio, Staveley-Smith,
  Ekers, Bowman, Briggs, Cappallo, Deshpande, Greenhill, Hazelton, Kaplan,
  Lonsdale, McWhirter, Mitchell, Morales, Morgan, Oberoi, Ord, Prabu, Shankar,
  Srivani, Subrahmanyan, Tingay, Webster, Williams, Williams, Kapi{\'{n}}ska,
  Morgan, Offringa, Wayth, Wu, Zheng, Murphy, Bell, Dwarakanath, For, Gaensler,
  Johnston-Hollitt, Lenc, Procopio, Staveley-Smith, Ekers, Bowman, Briggs,
  Cappallo, Deshpande, Greenhill, Hazelton, Kaplan, Lonsdale, McWhirter,
  Mitchell, Morales, Morgan, Oberoi, Ord, Prabu, {Udaya Shankar}, Srivani,
  Subrahmanyan, Tingay, Webster, Williams, \& Williams}]{Hurley-Walker2017}
Hurley-Walker, N., Callingham, J.~R., Hancock, P.~J., {et~al.} 2017, MNRAS,
  464, 1146

\bibitem[{Intema {et~al.}(2009)Intema, van~der Tol, Cotton, Cohen, van Bemmel,
  \& R{\"{o}}ttgering}]{Intema2009}
Intema, H.~T., van~der Tol, S., Cotton, W.~D., {et~al.} 2009, A{\&}A, 501, 1185

\bibitem[{Kaplan {et~al.}(2015)Kaplan, Tingay, Manoharan, Macquart, Hancock,
  Morgan, Mitchell, Ekers, Wayth, Trott, Murphy, Oberoi, Cairns, Feng,
  Kudryavtseva, Bernardi, Bowman, Briggs, Cappallo, Deshpande, Gaensler,
  Greenhill, {Hurley Walker}, Hazelton, {Johnston Hollitt}, Lonsdale,
  McWhirter, Morales, Morgan, Ord, Prabu, {Udaya Shankar}, Srivani,
  Subrahmanyan, Webster, Williams, \& Williams}]{Kaplan2015}
Kaplan, D.~L., Tingay, S.~J., Manoharan, P.~K., {et~al.} 2015, Astrophys. J.
  Lett., 809

\bibitem[{Kintner {et~al.}(2007)Kintner, Ledvina, \& {De Paula}}]{Kintner2007}
Kintner, P.~M., Ledvina, B.~M., \& {De Paula}, E.~R. 2007, Sp. Weather, 5,
  S09003

\bibitem[{Lane {et~al.}(2014)Lane, Cotton, van Velzen, Clarke, Kassim,
  Helmboldt, Lazio, \& Cohen}]{Lane2014}
Lane, W.~M., Cotton, W.~D., van Velzen, S., {et~al.} 2014, MNRAS, 440, 327

\bibitem[{Loi {et~al.}(2015{\natexlab{a}})Loi, Murphy, Cairns, Menk, Waters,
  Erickson, Trott, Hurley-Walker, Morgan, Lenc, Offringa, Bell, Ekers,
  Gaensler, Lonsdale, Feng, Hancock, Kaplan, Bernardi, Bowman, Briggs,
  Cappallo, Deshpande, Greenhill, Hazelton, Johnston-Hollitt, McWhirter,
  Mitchell, Morales, Morgan, Oberoi, Ord, Prabu, Shankar, Srivani,
  Subrahmanyan, Tingay, Wayth, Webster, Williams, \& Williams}]{Loi2015b}
Loi, S.~T., Murphy, T., Cairns, I.~H., {et~al.} 2015{\natexlab{a}}, Geophys.
  Res. Lett., 42, 3707

\bibitem[{Loi {et~al.}(2015{\natexlab{b}})Loi, Trott, Murphy, Cairns, Bell,
  Hurley-Walker, Morgan, Lenc, Offringa, Feng, Hancock, Kaplan, Kudryavtseva,
  Bernardi, Bowman, Briggs, Cappallo, Corey, Deshpande, Emrich, Gaensler,
  Goeke, Greenhill, Hazelton, Johnston-Hollitt, Kasper, Kratzenberg, Lonsdale,
  Lynch, McWhirter, Mitchell, Morales, Morgan, Oberoi, Ord, Prabu, Rogers,
  Roshi, Shankar, Srivani, Subrahmanyan, Tingay, Waterson, Wayth, Webster,
  Whitney, Williams, \& Williams}]{Loi2015a}
Loi, S.~T., Trott, C.~M., Murphy, T., {et~al.} 2015{\natexlab{b}}, Radio Sci.,
  50, 574

\bibitem[{Mangum \& Wallace(2015)}]{Mangum2014}
Mangum, J.~G. \& Wallace, P. 2015, Publ. Astron. Soc. Pacific, 127, 74

\bibitem[{McKay-Bukowski {et~al.}(2015)McKay-Bukowski, Vierinen, Virtanen,
  Fallows, Postila, Ulich, Wucknitz, Brentjens, Ebbendorf, Enell, Gerbers,
  Grit, Gruppen, Kero, Iinatti, Lehtinen, Meulman, Norden, Orispaa, Raita, {De
  Reijer}, Roininen, Schoenmakers, Stuurwold, \& Turunen}]{McKay-Bukowski2015}
McKay-Bukowski, D., Vierinen, J., Virtanen, I.~I., {et~al.} 2015, {KAIRA: The
  Kilpisj{\"{a}}rvi atmospheric imaging receiver array - System overview and
  first results}

\bibitem[{Mevius {et~al.}(2016)Mevius, van~der Tol, Pandey, Vedantham,
  Brentjens, de~Bruyn, Abdalla, Asad, Bregman, Brouw, Bus, Chapman, Ciardi,
  Fernandez, Ghosh, Harker, Iliev, Jeli{\'{c}}, Kazemi, Koopmans, Noordam,
  Offringa, Patil, van Weeren, Wijnholds, Yatawatta, \& Zaroubi}]{Mevius2016}
Mevius, M., van~der Tol, S., Pandey, V.~N., {et~al.} 2016, Radio Sci., 51, 927

\bibitem[{Narayan(1992)}]{Narayan1992}
Narayan, R. 1992, Philos. Trans. R. Soc. A Math. Phys. Eng. Sci., 341, 151

\bibitem[{Noordam(2004)}]{Noordam2004}
Noordam, J.~E. 2004, in Proc. SPIE, Vol. 5489 (SPIE), 817--825

\bibitem[{Or{\'{u}}s {et~al.}(2005)Or{\'{u}}s, Hern{\'{a}}ndez-Pajares, Juan,
  \& Sanz}]{Orus2005}
Or{\'{u}}s, R., Hern{\'{a}}ndez-Pajares, M., Juan, J.~M., \& Sanz, J. 2005, J.
  Atmos. Solar-Terrestrial Phys., 67, 1598

\bibitem[{Petit \& Luzum(2010)}]{Petit2010}
Petit, G. \& Luzum, B. 2010, {IERS Conventions (2010)}, ed. {Frankfurt am Main:
  Verlag des Bundesamts f{\"{u}}r Kartographie und Geod{\"{a}}sie}, IERS
  Technical Note 36

\bibitem[{Rees(1990)}]{Rees1990}
Rees, N. 1990, MNRAS, 244, 233

\bibitem[{Schaer(1999)}]{Schaer1999}
Schaer, S. 1999, {Mapping and Predicting the Earth's Ionosphere Using the
  Global Positioning System}, Vol.~59

\bibitem[{Tasse(2014)}]{Tasse2014}
Tasse, C. 2014, Astron. Astrophys., 566, A127

\bibitem[{Tasse {et~al.}(2017)Tasse, Hugo, Mirmont, Smirnov, Atemkeng, Bester,
  Hardcastle, Lakhoo, Perkins, \& Shimwell}]{Tasse2017}
Tasse, C., Hugo, B., Mirmont, M., {et~al.} 2017, eprint arXiv:1712.02078

\bibitem[{Tingay {et~al.}(2013)Tingay, Goeke, Bowman, Emrich, Ord, Mitchell,
  Morales, Booler, Crosse, Wayth, Lonsdale, Tremblay, Pallot, Colegate,
  Wicenec, Kudryavtseva, Arcus, Barnes, Bernardi, Briggs, Burns, Bunton,
  Cappallo, Corey, Deshpande, Desouza, Gaensler, Greenhill, Hall, Hazelton,
  Herne, Hewitt, Johnston-Hollitt, Kaplan, Kasper, Kincaid, Koenig,
  Kratzenberg, Lynch, McKinley, McWhirter, Morgan, Oberoi, Pathikulangara,
  Prabu, Remillard, Rogers, Roshi, Salah, Sault, Udaya-Shankar, Schlagenhaufer,
  Srivani, Stevens, Subrahmanyan, Waterson, Webster, Whitney, Williams,
  Williams, \& Wyithe}]{Tingay2012}
Tingay, S.~J., Goeke, R., Bowman, J.~D., {et~al.} 2013, Publ. Astron. Soc.
  Aust., 30

\bibitem[{van Haarlem {et~al.}(2013)van Haarlem, Wise, Gunst, Heald, McKean,
  Hessels, de~Bruyn, Nijboer, Swinbank, Fallows, Brentjens, Nelles, Beck,
  Falcke, Fender, H{\"{o}}randel, Koopmans, Mann, Miley, R{\"{o}}ttgering,
  Stappers, Wijers, Zaroubi, van~den Akker, Alexov, Anderson, Anderson, van
  Ardenne, Arts, Asgekar, Avruch, Batejat, B{\"{a}}hren, Bell, Bell, van
  Bemmel, Bennema, Bentum, Bernardi, Best, B{\^{i}}rzan, Bonafede,
  a.~J.~Boonstra, Braun, Bregman, Breitling, van~de Brink, Broderick, Broekema,
  Brouw, Br{\"{u}}ggen, Butcher, van Cappellen, Ciardi, Coenen, Conway, Coolen,
  Corstanje, Damstra, Davies, Deller, Dettmar, van Diepen, Dijkstra, Donker,
  Doorduin, Dromer, Drost, van Duin, Eisl{\"{o}}ffel, van Enst, Ferrari,
  Frieswijk, Gankema, Garrett, de~Gasperin, Gerbers, de~Geus, Grie{\ss}meier,
  Grit, Gruppen, Hamaker, Hassall, Hoeft, Holties, Horneffer, van~der Horst,
  van Houwelingen, Huijgen, Iacobelli, Intema, Jackson, Jelic, de~Jong, Juette,
  Kant, Karastergiou, Koers, Kollen, Kondratiev, Kooistra, Koopman, Koster,
  Kuniyoshi, Kramer, Kuper, Lambropoulos, Law, van Leeuwen, Lemaitre, Loose,
  Maat, Macario, Markoff, Masters, McFadden, McKay-Bukowski, Meijering,
  Meulman, Mevius, Middelberg, Millenaar, Miller-Jones, Mohan, Mol, Morawietz,
  Morganti, Mulcahy, Mulder, Munk, Nieuwenhuis, van Nieuwpoort, Noordam,
  Norden, Noutsos, Offringa, Olofsson, Omar, Orr{\'{u}}, Overeem, Paas,
  Pandey-Pommier, Pandey, Pizzo, Polatidis, Rafferty, Rawlings, Reich,
  de~Reijer, Reitsma, Renting, Riemers, Rol, Romein, Roosjen, Ruiter, Scaife,
  van~der Schaaf, Scheers, Schellart, Schoenmakers, Schoonderbeek, Serylak,
  Shulevski, Sluman, Smirnov, Sobey, Spreeuw, Steinmetz, Sterks, Stiepel,
  Stuurwold, Tagger, Tang, Tasse, Thomas, Thoudam, Toribio, van~der Tol, Usov,
  van Veelen, van~der Veen, ter Veen, Verbiest, Vermeulen, Vermaas, Vocks,
  Vogt, de~Vos, van~der Wal, van Weeren, Weggemans, Weltevrede, White,
  Wijnholds, Wilhelmsson, Wucknitz, Yatawatta, Zarka, Zensus, \& van
  Zwieten}]{vanHaarlem2013}
van Haarlem, M.~P., Wise, M.~W., Gunst, a.~W., {et~al.} 2013, A{\&}A, 556, A2

\bibitem[{van Weeren {et~al.}(2016)van Weeren, Williams, Hardcastle, Shimwell,
  Rafferty, Sabater, Heald, Sridhar, Dijkema, Brunetti, Br{\"{u}}ggen,
  Andrade-Santos, Ogrean, R{\"{o}}ttgering, Dawson, Forman, de~Gasperin, Jones,
  Miley, Rudnick, Sarazin, Bonafede, Best, B{\^{i}}rzan, Cassano, Chy{\.{z}}y,
  Croston, Ensslin, Ferrari, Hoeft, Horellou, Jarvis, Kraft, Mevius, Intema,
  Murray, Orr{\'{u}}, Pizzo, Simionescu, Stroe, van~der Tol, \&
  White}]{vanWeeren2016b}
van Weeren, R.~J., Williams, W.~L., Hardcastle, M.~J., {et~al.} 2016,
  Astrophys. J. Suppl. Ser., 223, id2

\end{thebibliography}



\end{document}